\documentstyle[aps,pra,psfig,twocolumn,floats]{revtex}

\newcommand{\QKD}{{\sc qkd}} 
\newcommand{\PDC}{{\sc pdc}} 
\newcommand{\WCP}{{\sc wcp}} 

\begin{document}
\draft
\title{Security against individual attacks for realistic quantum key distribution}
\author{Norbert L\"utkenhaus}
\address{Helsinki Institute of Physics, PL 9, FIN-00014
Helsingin yliopisto, Finland}
\date{\today}
\maketitle
\begin{abstract}
I prove the security of quantum key distribution  against
individual attacks for realistic signals sources, including weak
coherent pulses and downconversion sources. The proof applies to the
BB84 protocol with the standard detection scheme (no strong reference
pulse). I obtain a formula for the secure bit rate per time slot of an
experimental setup which can be used to optimize the performance of
existing schemes for the considered scenario. 
\end{abstract}
\narrowtext

\section{Introduction}
The first complete protocol for quantum key distribution (\QKD) has been
introduced by Bennett and Brassard in 1984 \cite{bennett84a} following
earlier ideas by Wiesner \cite{wiesner83a}. Since then,
this protocol (BB84 for short) has
been implemented by several groups
\cite{marand95a,townsend98a,breguet94a,muller97a,zbinden98a,ribordysub99a,buttler98a,hughessub99a,bourennane99a,franson94a,dusek99b}. For an overview
containing more details about the background, the experimental
implementation and the classical evaluation procedure see for
example \cite{zbinden98a,bennett92a,phoenix93a,brusssub98b}. 

The basic idea of the BB84 protocol is to use a random string of
signal states which, for example, can be realized as single photons
in horizontal, vertical, right circular, or left circular
polarization states. These are two set of states which are orthogonal within
each set, and have overlap probability $1/2$ between the sets. If the
receiver chooses at random between a polarization analyzer
for  linear polarization and one for circular polarization, then they
obtain in this way a {\em raw key} \cite{huttner94a}. From this they
distill the {\em sifted key} by
publicly exchanging information about the polarization basis of the
signals and the measurement apparatus. They keep only those bits where
the basis is the same for the signal and the measurement, since those
signals give a deterministic relation between signal and measurement
outcome.

The practical implementations deviate from the theoretical abstraction
used in the original proposal in two important points. The first is
that  the signal
states do not have the correct overlap probabilities. Especially in
the photonic realization, the signals contain contributions from higher
photon numbers and from the vacuum state which cause this deviation. The second point is that the quantum channel in these
implementations (optical fibers) shows a considerable loss. It has been
shown earlier \cite{huttner95a,yuen96a} that the combination of the two
effects open up a security gap. The extent of this security gap has
been extensively illuminated for different signal sources in
\cite{brassardsub99a} giving necessary conditions  on the feasibility of \QKD\
without restriction to any particular class of eavesdropping attacks. From these
results one can conclude that most current experiments are performed in
a parameter regime where the necessary conditions for security are violated.

In the present  work I will complement these results by a positive proof
of security for a scenario where the power of the eavesdropper is
restricted to attacking signals separately (individual attack). This
restriction allows us to prove the security for a realistic protocol,
i.~e.~one where all components are known and work efficiently. 

It is necessary to distinguish this work from earlier work by other
groups. Lo and Chau \cite{lo99a} gave a proof of principle for the
security of quantum key distribution. At present, it is not possible
to use their proof to implement secure \QKD\ since the procedure involves
devices to manipulate qubits coherently in order to allow
fault-tolerant computing. The approach of Mayers \cite{mayers98a} is
certainly the most advanced result towards practical \QKD\ which is
provably 
secure against all eavesdropping attacks on the signals. However, the
proof assumes ideal single photon signals, and, at
present, we do not have an extension of that proof which can cope with
realistic signal sources and  effective error correction codes,
although work in these directions is in progress. 

The restriction to eavesdropping on individual signals allows a much
simpler analysis of a realistic scenario, and it is therefore
advisable to use this scenario as a study for the generalization in
the sense of Mayer's proof. Furthermore, the results are interesting
in their own right: it seems to be impossible to perform collective
measurements on the signals with today's technology. Therefore, \QKD\
secure against individual attack will today create keys which are
secure against future developments in coherent eavesdropping
strategies, since tomorrows technology cannot be used for todays
eavesdropping strategy. This is in contrast to the implication of an
increase of future computation power or improvements in algorithms
which threatens todays use of classical encryption schemes.

In this paper I will derive a formula for the gain of secure bits per
signal sent, that is per time slot of the experiment. These formulas
are presented only in the limit of long keys, so that the influence of
the necessary authentication of the key and all statistical influences
regarding the number of errors etc.~can be neglected. It is necessary
to embed these results into a full protocol, derived, for example, in
\cite{hughessub99a,nl99a,slutsky98a} to which I refer the reader for further details. 

This paper is organized as follows. In Sec.~\ref{singlephoton} I
will introduce  the essential elements of practical quantum cryptography
and report the relevant findings for single photon signals. These
results are then extended in Sec.~\ref{multiphoton} to signal
sources which generate the signal states by rotating a state in one
polarization to that of the ideal BB84
polarizations. In
Sec.~\ref{exploring}, the
resulting gain formula is explored for two choices for the signal
source, namely weak coherent pulses (\WCP) and parametric
downconversion (\PDC). The results are discussed
in Sec.~\ref{discussion}.

\section{Security against individual attacks for single photon
sources}
\label{singlephoton}
To investigate the security of \QKD\ one needs to investigate the
trade-off between the information gathered by the eavesdropper and the
amount of disturbance caused thereby. The trade-off between the Shannon mutual
information and the bit error rate in the sifted key has been
investigated by several author for restricted attacks
\cite{huttner94a,ekert94a} and for  the general
individual attack \cite{fuchs97a}. The results show that the gathered
Shannon information for the typically observed 
error rate of about $1$--$5$\% is too high to allow the sifted key to be used directly for
cryptographic purposes. However, we can first correct the errors and
then apply the technique of {\em generalized privacy amplification}
\cite{bennett95a} to
distill from the sifted key a new shorter key, which fulfills the
security requirements. These techniques are purely classical. Both
steps, the error correction and the privacy amplification, will reduce
the number of gained secure bits.

\subsection{Error correction}
Error correction is performed by the exchange of redundant information
about the key, e.g. in form of parity bits, via the public channel. Since Eve has access to the
public channel, we have to take care of this flow of
side-information. This can be done by using a short initial shared
secret key to encrypt the parity bits in a one-time pad method. Note that in practice we cannot realize any public channel
which is safe against tampering by Eve by technology alone. Therefore,
sender and receiver need to
share a secret key anyway to overcome this problem by the classical
method of authentication \cite{carter79a,wegman81a}. As a consequence
of this method of control of the side-information, we need to know how
many bits need to be encrypted, which is equivalent to the number of
exchanged parity bits.

It is clear, that one has to be careful to implement an efficient
error correction protocol, since we have to regain at least the number
of secret bits used for the encryption of the parity bits. The ratio
between minimum
number of redundant bits $N^{\rm Shannon}_{\rm corr}$ needed to correct a
key of length $n$  is given according to
Shannon \cite{shannon48a} by
\begin{equation}
\frac{N^{\rm Shannon}_{\rm corr}}{n} = -e \log_2 e- (1-e) \log_2
(1-e)\; ,
\end{equation}
where $e$ is the observed error rate in the sifted key. In this limit
the probability that the errors can be corrected can come arbitrarily
close to unity. However,
Shannon's proof of the existence of error correction codes reaching
this limit is not constructive, and the limit is obtained only by  large codes. These are not easily implemented because of the
required computational resources. We have therefore to search for error correction
tools which come close to this limit. As discussed in \cite{nl99a}, it
is hard even to approach the Shannon limit with error correction codes
which use uni-directional classical communication only. Fortunately, a
more efficient bi-directional code exists \cite{brassard93a},
which uses $f[e]\;  N^{\rm Shannon}_{\rm corr}$ bits for error correction
with a correction factor $f[e]$ listed in table  \ref{brassardsalvail}.
\begin{table}[htb]
\caption{\label{brassardsalvail} Example of the performance of the
bi-directional error reconciliation protocol by Brassard and Salvail
\protect\cite{brassard93a}. The values are taken from that paper. Here
$e$ is the observed error rate, while $f[e]$ is the ratio of actually
needed redundant bits to the corresponding number of the Shannon
limit. (I used the upper bounds for $I(4)$ provided in the
reference.)}
\begin{tabular}{ll}
$e$ & $f[e]$\\\hline
0.01 & 1.16\\
0.05 & 1.16 \\
0.1 & 1.22 \\
0.15 & 1.35 \\
\end{tabular}
\end{table}

\subsection{Generalized privacy amplification}
In this section I report on the fraction $\tau_1$ of bits by which we
need to shorten the sifted key so that we obtain a secure key.
The aim of \QKD\ is to obtain a secure key in the sense that Eve has
no information on that key. This can be made precise by two
properties: 1) a key $x$ of length $n_{\rm final}$ should have equal a
priori probability $p(x)=2^{-n_{\rm final}}$ and 2) the difference
between the a priori and a posteriori probability, as measured by the
Shannon information, should vanish. These two properties can be
summarized in the demand that the expected Shannon entropy $H[\langle p(x|M)\rangle_M]$
of the a posteriori probability distribution $\langle p(x|M)\rangle_M$,
after Eve's gathering of measurement results and classical
communication $M$, should approach $n_{\rm final}$. (Here $\langle
\dots \rangle_M$ denotes the expectation value with respect to the
measurement outcome $M$.) Generalized
privacy amplification \cite{bennett95a} achieves that by hashing the
corrected sifted key into a shorter key by {\em hash functions}
\cite{carter79a,wegman81a} such that we obtain the bound \cite{bennett95a} (see
\cite{nl99a} for the extension to the expectation values with respect to $M$)
\begin{equation}
H[\langle p(x|M)\rangle_M] \geq n_{\rm final}- \log_2\left(2^{n_{\rm final}} \langle
p_c[p(x|M)]\rangle_M + 1 \right) \; .
\end{equation}
Here $p_c[p(x|M)]$ is a measure of the a posteriori
probability on the corrected sifted key $x$ of length
$n_{\rm sif}$.
This measure is the collision probability, defined as
\begin{equation}
p_c[p(x|M)] = \sum_{x}p^2(x|M)  \; .
\end{equation}
If we choose the length of the final key to be
\begin{equation}
n_{\rm final}=n_{\rm sif}(1-\tau_1)-n_{\rm S} \; ,
\end{equation}
the estimate becomes, after a further simplifying estimation \cite{bennett95a},
\begin{equation}
H[\langle p(x|M)\rangle_M] \geq n_{\rm final}-\frac{2^{-n_{\rm S}}}{\ln 2}
\end{equation}
with 
\begin{equation}
\tau_1 = 1 + \frac{1}{n_{\rm sif}} \log_2\langle
p_c[p(x|M)]\rangle_M \; .
\end{equation}
Clearly, we can approximate an ideal secret key arbitrarily close by the
choice of the security parameter $n_{\rm S}$. For long keys, only the
shortening fraction $\tau_1$ needs to be taken account of. 

The above formulas show that an upper bound on the expected collision
probability leads to a lower bound on the Shannon information. Such
bounds have been provided for the BB84 protocol in
\cite{nl99a,slutsky98a,nl96a} for various scenarios. We concentrate here on the
case that the errors in the sifted key are corrected (as opposed to
discarding the corresponding bits) using the bi-directional error
correction procedures. 
We define 
the collision probability $p_c^{(1)}(e)$, as a function of the error
rate $e$ in the sifted key, for a single bit of the corrected sifted key implicitly by  $\langle
p_c[p(\overline{x}|M)]\rangle_M=\left(p_c^{(1)}[e] \right)^{n_{\rm
sif}}$ and find  the bound \cite{nl99a}
\begin{equation}
\label{pcsingle} 
p_c^{(1)}(e) \leq \left\{ \begin{array}{ll} 
\frac{1}{2} + 2 e - 2 e^2 & \mbox{for 
$e\leq 1/2$} \\ 1 & \mbox{for $1/2 \leq 
e$} \end{array} \right.  \; .
\end{equation} 
which gives, finally, 
\begin{equation} 
\label{taucorrleak} \tau_1(e) \leq 
\left\{\begin{array}{ll}\log_2 \left( 1 + 4 e - 4 
e^2 \right) & \mbox{for $e \leq 
1/2$} \\ 1 & \mbox{for $ 1/2 \leq e $} \end{array} 
\right.  \; .
\end{equation} 
The estimate is valid for uni-directional protocols as well since the
additional information flow to Eve during bi-directional error
correction takes, apparently, the form of a {\em spoiling information}
in the sense of \cite{bennett95a}. As
pointed out in \cite{nl99a}, we have to be careful in dealing with
ambiguous detections, for example clicks in both detectors monitoring
orthogonal polarizations. A way to deal with that is to randomly
assign a bit value to those events. Discarding those events would open
a loophole for the eavesdropper.

\subsection{Gain formula for single photon signals}
\begin{figure}[htb]
\centerline{\psfig{width=8cm,file=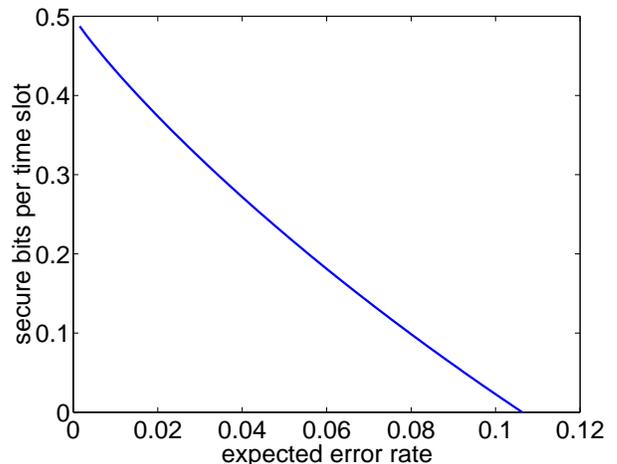}} 
\caption{\label{singlephotongain} Gain of secure bits per time slot as a
function of the observed error rate $e$ for an ideal channel for
single-photon signals and ideal
error correction. }
\end{figure}

We can summarize the effects of error correction and privacy
amplification by a gain formula for the limit of long keys. It is
given by
\begin{eqnarray}
\lefteqn{G^{\rm single}}& & \\
& & = \frac{1}{2} p_{\rm exp} \left\{1 -\tau_1 + f[e] \left(e \log_2
e+ (1-e) \log_2 (1-e)\right)\right\} \; . \nonumber
\end{eqnarray}
Bob's detector is triggered with probability $p_{\rm exp}$, taking
into account channel losses and imperfect detection efficiencies, and in
half of the cases the signal is entered into the sifted key. From the
length of the sifted key we have to deduct the cost of error
correction and of privacy amplification.
The resulting rate for a lossless transmission, $p_{\rm exp} =1$, and
ideal error correction, $f[e]=1$, is shown in figure
\ref{singlephotongain}. 
From there it becomes clear that the maximal tolerated error rate for
this approach is around $11 \%$. 

\section{Extension to multi-photon sources with ideal polarizations} 
\label{multiphoton}
To generalize the results of the previous section to realistic signal
sources we first need to consider which signals states we can
generate. We find that the typical sources show a simple structure which
allows us to describe the optimal eavesdropping strategy. As a
consequence, we can bound Eve's collision probability using the
results derived for single photon signals.

\subsection{Realistic signal sources}
The signal sources described here generate the signal from some state
in one polarization mode by changing its polarization to one of the
four BB84 polarization modes.

Typically, there will be no fixed relation between the optical phase
of subsequent signals. As a result, Eve ``sees'' the phase averaged
form of the signals \cite{brassardsub99a}  which take the form of a mixture of Fock states in the chosen
polarization mode. (The off-diagonal terms average out to zero.) This
observation, in fact, simplifies the analysis of security.

It should be noted that even if the source should bear some phase
relation between subsequent pulses, this relation can be destroyed by
including a phase randomizer which selects at random an optical phase
for each signal. This is needed, for example, for the ``plug and play
scheme'' by the Geneva group \cite{muller97a}. Note that the so-called
phase encoding \cite{marand95a} is basically equivalent to the the
polarization encoding. This is so because  the four BB84 polarizations
can be expressed,
mathematically,  as a relative
phase between two modes. Phase encoding uses the relative phase
between two spatially separated modes (in the same fiber and the same
polarization mode). They are therefore equivalent. However, in some
implementations one of the spatial mode pulses has a bigger amplitude to
implement some kind of strong reference pulse for an interference in
Bob's detector, as proposed in the two state protocol
\cite{bennett92b} and the ``4+2'' protocol \cite{huttner95a}. The security analysis presented here does not apply to
these set-ups. 

\subsection{Estimation of the collision probability}
We have seen above that for the signal sources investigated here, the
signals are mixtures of Fock states in the chosen polarization
mode. It turns out that Eve can split the photon number of each signal
containing two or more photons by extracting  one or
more photons out of the signal  
such that both parts retain their original polarization. (See appendix
\ref{photonsplitting}.) This can be
achieved by interactions of the Jaynes-Cummings type which are
preceded by a quantum non-demolition measurement of the total photon
number of the signal. This stands not in contrast to the statement of
Yuen \cite{yuen96a} that it is not possible to extract a
photon from an arbitrary state, since here we are talking only about
states with known total photon number, and where all photons are in a
single, though unknown, mode. On the other hand, it is unclear what it
would mean for other states to extract a photon such that the
extracted photon and the remaining states have an unaltered
polarization. Eve can
perform a measurement on her photons after receiving the information
about the polarization basis of the signals, and she therefore will
know the bit-value of these signals. On the other hand, she does not
cause any errors on Bob's side, since the photons arrive there with
the original polarization. 

We can summarize this in the statement that the collision probability
on each bit in the sifted key which stems from a multi-photon signal
is equal to $1$, and all errors in the sifted key are due to eavesdropping on
single photon signals contribution to the sifted key.

The collision probability for the sifted key factorizes into the
product of collision probabilities for each bit. If we know an upper
bound on the number $m$ of multi-photon signals contributing to the sifted key, then
we can estimate the collision probability on the sifted key of length
$n_{\rm sif}$ by the single bit collision probabilities for single photon
signals $p_c^{(1)}$ and that for multi-photon signals $p_c^{(m)}=1$ as
\begin{equation} 
    p_{c} \leq \left(p_{c}^{(m)}\right)^m 
    \left(p_{c}^{(1)}\right)^{n_{\rm sif}-m} = 
    \left(p_{c}^{(1)}\right)^{n_{\rm sif}-m}\; .
\end{equation} 
The value of the error rate at which $p_{c}^{(1)}$ from
Eq. \ref{pcsingle} is evaluated, has to be rescaled since all errors
are assumed to stem from eavesdropping on the single-photon
signals. We therefore find
\begin{equation} 
    p_{c} \leq \
     \left(p_{c}^{(1)}\left[e \frac{n_{\rm sif}}{n_{\rm sif}-m}\right]\right)^{n_{\rm sif}-m}\; ,
\end{equation} 
 which gives the fraction of the key which has to be discarded during
 privacy amplification as
\begin{equation} 
    \label{taum} 
\tau_1^{(m)}(e^{(1)}) = 1 + \frac{n_{\rm sif}-m}{n_{\rm sif}} 
\log_2 p_c^{(1)}\left[e \frac{n_{\rm sif}}{n_{\rm sif}-m} \right]\; .
\end{equation} 
The number of multi-photon bits contributing to the sifted key can be
bounded once we know the source characteristic in the form of
probabilities $S_0$, $S_1$, and $S_m$ for the signal to contain zero,
one, or more than one photon. Eve will use all multi-photon signals
while she suppresses partly single-photon signal to obtain the desired
fraction $p_{\rm exp}$ of signals successfully detected by Bob. Therefore
the expectation value for the number  $m$ of signals stemming from
multi-photon signals is given by $\langle m \rangle = S_m n_{\rm
tot}$, where $n_{\rm tot}$ is the total number of signals sent by Alice.  We can use a theorem by Hoeffding \cite{hoeffding63a} to
relate the expected number of multi-photon signals $\langle m \rangle$ to
the actually created number of such signals $m$ for a key of length
$n_{\rm sif}$ with some probability.  The statement is that the
inequality
\begin{equation} 
    \left| \langle m \rangle - m\right| \leq \delta \; n_{\rm tot}
\end{equation}
for some chosen value of $\delta$ holds with a probability
$P>1-\exp\left( -2 n_{\rm tot} \delta^2\right)$.  This means, that we can choose
$m=\langle m \rangle$ since we deal in this article only with the limit of
large keys.  For experimental realizations, however, one has to keep
an eye on the choice of $\delta$ which might be rather small.  Then
$n_{\rm tot}$ has to be quite large to obtain a reasonable value for $P$.  More
discussion concerning the statistical issue can be found in
\cite{nl99a}.

\subsection{Gain formula for realistic signal sources}
The gain formula for the considered signal sources is now given by
\begin{eqnarray}
\label{prediction}
\lefteqn{G^{(\rm multi)} = \frac{1}{2} p_{\rm post}\; p_{\rm exp} 
  \bigg\{\frac{n_{\rm sif}-\langle m \rangle}{n_{\rm sif}} }& & \\
& \times & \left(1-\log_2 \left[1 + 4 e
\frac{n_{\rm sif}}{n_{\rm sif}-\langle m \rangle} - 4\left( e \frac{n_{\rm sif}}{n_{\rm sif}-\langle m \rangle} \right) ^2 \right]\right)  \nonumber\\
 &+ &  f[e] \left[ e \log_2 e + (1-e) \log_2 (1-e) \right] \bigg\}\; . \nonumber
\end{eqnarray}
Here I included a factor $p_{\rm post}$ as the post-selection
probability of the signal. We need this for a consistent presentation
of the results using parametric downconversion, since there Alice
performs a post-selection for each time slot. The quantities  $p_{\rm
exp}$, $n_{\rm tot}$,
and $S_0$, $S_1$, and $S_m$ refer always to the post-selected signals
to emphasise the view that post-selection is the state preparation.
All parameters needed to evaluate this expression are
actually observables of the experiment. The value of $n_{\rm sif}$ is
agreed between Alice and Bob, the value of $n_{\rm tot}$ becomes known
to them during the key generation and leads to $p_{\rm exp} =
\frac{n_{\rm sif}}{n_{\rm tot}}$. The value of $e$ are
directly observed. The value of $S_m$ is indirectly measurable in Alice's
laboratory and leads to $\langle  m \rangle =
S_m n_{\rm tot}$. We can reformulate the expression for the gain as
\begin{eqnarray}
    \label{gainrate}
\lefteqn{G^{(\rm multi)}= \frac{1}{2} p_{\rm post}\; p_{\rm exp} \bigg\{\frac{p_{\rm
exp}-S_m}{p_{\rm exp}}}& &  \\
  & & \times 
    \left(1-\log_2 \left[1  + 4 e 
    \frac{p_{\rm exp}}{p_{\rm exp}-S_m}-4\left(e 
    \frac{p_{\rm exp}}{p_{\rm exp}-S_m}\right)^2\right]\right)   \nonumber \\ 
    & &  + f[e] \left[e\log_2 
    e + \left(1-e\right) \log_2 
    \left(1-e\right)\right]\bigg\} \nonumber
\end{eqnarray}
so that it is expressed entirely in measurable quantities. In this
form we can use it to estimate the gain for a running experiment
without having to implement the classical procedures of error
correction and privacy amplification. 

\section{Simulation for experiments}
\label{exploring}
To simulate the gain we can obtain  from an experimental set-up, we
need to model the photon number distribution of the source in more detail. Here we
need more than the three probabilities $S_0$, $S_1$, and $S_m$ since
the probability $p_{\rm exp}$ depends on the photon number distribution
within the multi-photon signals as well. Furthermore, we need to model
the expected error rate of the experiment. 

In my calculation I take account of the photon number distribution of
the signal source and the losses in the quantum channel. Bob's
detection unit varies in different set-ups by the number of detectors
etc. The parameters entering the calculation here are the
single-photon detection efficiency $\eta_{\rm B}$ and the dark count rate
$d_{\rm B}$, both given for 
the whole detection unit. The dark count rate  is measured as dark
count detections per time slot, i.e. gating window.

\subsection{General formulas}
The probability $p_{\rm exp}$ that Bob detects a signal has two
sources, one coming from the detection of signal photons $p_{\rm exp}^{\rm signal}$, the other
from the dark counts of the detectors $p_{\rm exp}^{dark}$. The
combination gives
\begin{equation}
\label{pexp}
  p_{\rm exp} = p_{\rm exp}^{\rm signal} + p_{\rm exp}^{dark}-  p_{\rm exp}^{\rm signal}
  p_{\rm exp}^{dark}
 \end{equation}
where I assume that the dark counts are independent of the signal
  photon detection. 
Let $S_i$ be the probability that the source sends $i$ photons, then
the probability that Bob's detector is triggered by a signal photon is
given as a function of the detection efficiency
$\eta_{\rm B}$ and a transmission efficiency of the channel $\eta_T$ by
\begin{equation}
\label{psignalexp}
p_{\rm exp}^{\rm signal} = \sum_{i=1}^\infty S_i \sum_{l=1}^i {i \choose l} \left(\eta_{\rm B}
\eta_T\right)^l \left(1- \eta_{\rm B} \eta_T\right)^{i-l} \; .
\end{equation} 
The dark count distribution is simply given by 
\begin{equation}
\label{pdarkexp}
p_{\rm exp}^{dark}=d_{\rm B} \; .
\end{equation}

The error rate stems, again, from two sources. The first is an error
rate for the detected signal photons, which is due to alignment errors
or fringe visibility. The probability of an error per time slot due to
this mechanism is modeled  by $p^{\rm error}_{\rm align} = c\;
p_{\rm exp}^{\rm signal}$ with a constant $c$.  The dark count contribution to the same error
probability is given by $p^{\rm error}_{\rm align} = \frac{1}{2} d_{\rm B}$ since a
dark count will result at random in  one of the two measurement results
for Bob, so that in half of the cases an error is created. Then the error
rate in the sifted key is modeled by
\begin{equation}
e \approx \frac{c\; p_{\rm exp}^{\rm signal}+\frac{1}{2} d_{\rm B}}{p_{\rm exp}} \; 
\end{equation}
in a regime where coincidences between dark counts and real counts can
be neglected.
For optical fibers, the losses in the quantum channel can be derived
from the loss coefficient $\alpha$ measured  in dB/km, the length of the fiber
$l$ in km and the loss in Bob's detection unit $L_c$ in dB as
\begin{equation}
\label{etatdef}
\eta_{\rm T} = 10^{-\frac{\alpha l + L_c}{10}} \; .
\end{equation}
Typical values for the fibre loss $\alpha$ in the three telecommunication windows at $0.8 \mu{\rm
m}$, $1.3 \mu{\rm m}$, and $1.5 \mu{\rm m}$ are $2.5$ dB/km, $0.35$
dB/km, and $0.2$ dB/km respectively.

\subsection{Weak coherent pulses}
In most experiments for \QKD\ the signal source is a strongly
attenuated laser pulse. 
The sources uses in typically experiments, e.g. laser diodes, emit
pulses which optical phases are set at random by the initiating spontaneous
emission. Therefore these sources fall into the category for which our
arguments apply.

The photon number is Poisson distributed with $S_i =
\exp(-\mu) \mu^i/i!$ and mean photon number $\mu$. Therefore we obtain
\begin{eqnarray}
S_m &=& 1-\left(1+ \mu\right)\exp(-\mu) \\
p_{\rm exp}^{\rm signal} &=& 1- \exp(-\eta_{\rm B} \eta_{\rm T} \mu)
\end{eqnarray}
which allow us together with the
Eq.~(\ref{gainrate}-\ref{etatdef}) and a
post-selection probability $p_{\rm post}=1$ to calculate
the expected gain per time slot of an experiment with weak coherent
pulses. 

\begin{figure}[htb]
\centerline{\psfig{width=8cm,file=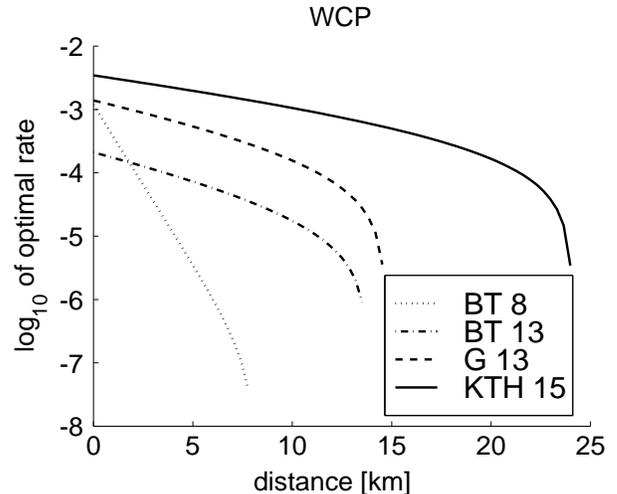}} 
\caption{\label{WCPgain} Weak coherent pulses: The rate of secure key bits per time slot for
realistic parameters described in the literature. (See table
II). The rate needs to be multiplied with the repetition
rate of the apparatus to obtain the true rate per second. Note that
the main effect for the shown experiments is the different absorption
rate of that fiber at the respective wavelength. Furthermore, these
experiments were not optimized with respect to the gain presented here.}
\end{figure}

We evaluate the resulting gain rate using parameter sets taken from
the literature. (See table II.)
\begin{table}[htb]
\label{parameters}
\caption{Parameters for quantum key distribution experiments taken from the literature.}
\begin{tabular}{ll|rrrr}
&  & BT 8 & BT 13 & G 13 & KTH 15\\
& & \cite{townsend98a}&\cite{marand95a}&\cite{ribordysub99a}&\cite{bourennane99a}\\
\noalign{\smallskip}\hline\noalign{\smallskip}
wavelength [nm]& & 830 &1300&  1300&1550\\
channel loss [dB/km] & $\alpha$& 2.5& 0.38 &0.32 & 0.2\\
receiver loss [dB]& $L_c$ & 8& 5 &  3.2 & 1\\
signal error rate [\%]&$e$& 1 & 0.8  & 0.14 & 1\\
dark counts [per slot]&$d_{\rm B}$&$ 5\times10^{-8}$ &$ 10^{-5}$ &$8.2\times 10^{-5}$&$ 2
\times 10^{-4}$\\
detection efficiency [\%]&$\eta_{\rm B}$& 50 & 11 & 17 & 18
\end{tabular}
\end{table} 
When we keep all parameters fixed and vary the
expected photon number of the signal, we obtain a gain curve with a
clear maximum. Furthermore, if the the photon number is too low, we
cannot obtain a positive gain because of the dark count rate of Bob's
detector. On the other hand, for large photon numbers we cannot obtain
a positive gain because of the high multi-photon probability for the
signals. We concentrate on the optimal choice of the expected photon
number which yields the maximal gain rate. Now we can vary the length
of the transmission line. The resulting graphs are shown in figure
\ref{WCPgain}.
 We see that the gain rate drops roughly exponentially with
the length of the transmission before it starts to drop faster due to
the increasing influence of the dark counts. The initial behavior is
mainly due to the multi-photon component of the signals while the
influence of the error-correction part is small. 
In this regime we can bound 
the gain by the approximation
\begin{eqnarray}
\label{glimit}
G &\leq & \frac{1}{2}\left( p_{\rm exp} - S_m\right) \\
& = & \frac{1}{2}\left\{(1+\mu) \exp(-\mu) - \exp ( -\eta_{\rm B} \eta_{\rm T} \mu)
\right\} \; .
\end{eqnarray}
This expression is optimized if we choose $\mu = \mu_{\rm optm}$ which
fulfills
\begin{equation}
\eta_{\rm B} \; \eta_{\rm T} \exp( -\eta_{\rm B} \eta_{\rm T} \mu_{\rm optm})-\mu_{\rm optm}
\exp(-\mu_{\rm optm}) = 0 \; .
\end{equation}
Since for a realistic setup we expect that $\eta_{\rm B} \eta_{\rm T} \ll 1$, we
find $\mu_{\rm optm} \approx \eta_{\rm B} \eta_T$. In this approximation we
find the approximate upper bound
\begin{equation}
\label{Gapprox}
G \approx \frac{1}{4}\eta_{\rm B}^2 \; \eta_T^2 \; .
\end{equation}
As the distance increases and
the influence of the dark counts and the error correction grows, this
approximation is no longer valid. Instead, we find in the numerical
simulations that the optimal
photon number is even lower. Note that in the real experiments much
higher photon number have been used. Typically, these higher photon
numbers do not allow secure key distribution over the reported
distances. 

The approximate situation described above illuminates another
interesting feature. As noted in \cite{brassardsub99a}, technical
limitations on detectors limit the distance over which we can perform
secure \QKD\ with weak coherent pulses, and the presented security
proof is in accordance with it. This limit can be
stretched  as the technology improves. However, the obtained distance
is only one characteristic of a setup. Another is  the
obtained rate. 
We find that the gain rate per time slot is limited already
by the use of the Poissonian photon number distribution and the loss
in the optical fiber.

\begin{figure}[htb]
\centerline{\psfig{width=8cm,file=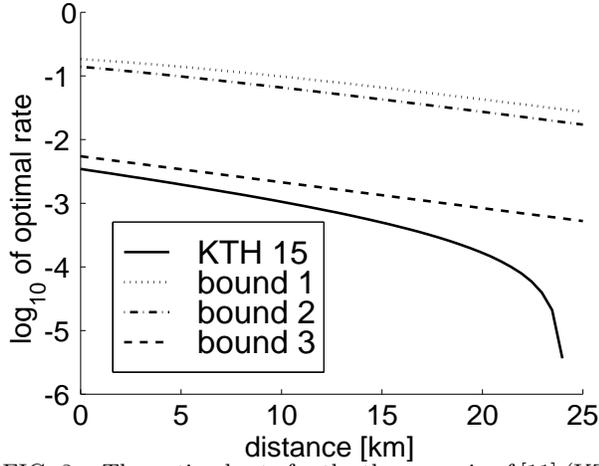}} 
\caption{\label{boundfigure} The optimal rate for the the scenario of
\protect\cite{bourennane99a} (KTH 15). Bound $1$ describes the optimal
possible rate given the use of Poissonian photon number distribution
and the loss of the quantum channel. Bound $2$ takes into account
additionally the given loss in Bob's receiver, while bound $3$ even
includes the detection inefficiency of Bob's detector. Therefore,
bound $3$ represents the approximation (\protect\ref{glimit}).  }
\end{figure}

We can evaluate Eqn.~(\ref{Gapprox}) for perfect detection devices and
get a bound 1 shown in Fig.~\ref{boundfigure} in the case of the KTH
set-up.
 The gap between bound 1  and the exact result shows how
much room is left for improvements of Bob's detection apparatus. The
bounds 2 and 3 take into account in addition to the fiber loss the
loss in Bob's detection device and the detection efficiency. We find
that bound 3 is already a good approximation to the exact results, at
least for short and medium distances. This shows that the multi-photon
aspect is for these distances the dominating effect compared to the
effect of error correction and the influence of eavesdropping on
single-photon signals, which are responsible for the gap between bound
3 and the exact curve. 
In order to compare the performance of different setups, one
would need to multiply the gain rate with the signal repetition rate
of the set-up to obtain the rate of secret bits per second. This
repetition rate may be vastly different for some applications, so that
the gain rate shown in Fig.\ref{WCPgain} is only a starting point in
optimizing the secure bit rate for a specific application. However, it
shows clearly the variation of the performance as the distance varies,
including the maximal possible distance.

\subsection{Parametric downconversion for triggering}
The results of the previous section illustrates that the coverable distance
for \QKD\ is limited. As shown explicitly in \cite{brassardsub99a},
this distance can be increased by the usage of other signal sources,
especially by the use of parametric downconversion. Note, however,
that it has been shown there that even perfect single photon sources will lead to a limited
coverable distance  due to Bob's dark count rate. 

\begin{figure}[htb]
\centerline{\psfig{width=8cm,file=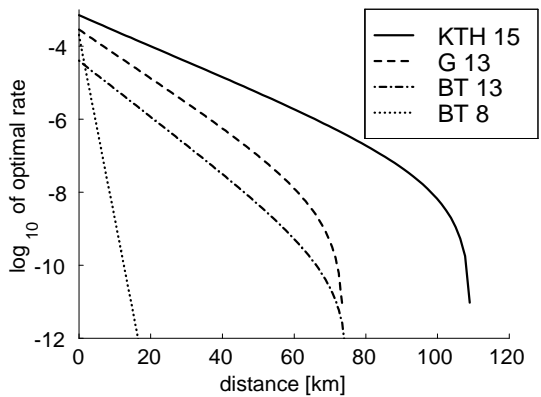}} 
\caption{\label{PDCgain} Parametric downconversion as triggering
device: The rate of secure key bits per time slot for
realistic parameters described in the literature. The triggering mode
is adapted to the $800$ nm detector of the BT experiment. The signal
mode is adapted to one of the four studied cases. (See table
II). The rate needs to be multiplied with the repetition
rate of the apparatus to obtain the true rate per second. }
\end{figure}

I will discuss here only the use of parametric downconversion (\PDC)
as a triggering mechanism, although more sophisticated techniques
using EPR states are possible. For that we consider the non-degenerate
parametric amplifier described by the parameter $\chi$ as the product of the
coupling constant and the interaction time of the process. This
creates the two-mode state \cite{walls94a}
\begin{equation}
|\Psi\rangle =(\cosh \chi)^{-1} \sum_{n=0}^\infty ( \tanh \chi)^n
|n,n\rangle \; .
\end{equation}
Alice monitors the first mode with a detector described by detection
efficiency $\eta_{\rm A}$ and dark count rate $d_{\rm A}$. Only coincidences
between Alice's and Bob's detector will be taken into account when
forming the sifted key. For a low dark count rate and a small
parameter $\chi$ (note that $\sinh^2 \chi$ is the expected photon
number in one mode) we can neglect coincidences between dark counts
and detection events and associate Alice's detection event with the
POM element 
\begin{equation}
E_{\rm click} = d_{\rm A} |0\rangle\langle 0| + \sum_{n=1}^{\infty}
(1-(1-\eta_{\rm A})^n) |n\rangle\langle n|\; .
\end{equation}
The signal state conditioned on Alice's detection event is then given by
\begin{eqnarray}
\rho_{\rm click} &=& \frac{1}{p_{\rm post}}{\rm Tr_A}\left( E_{\rm
click}|\Psi\rangle\langle \Psi|
\right)\\
 & =& \frac{1}{p_{\rm post}}  \frac{d_{\rm A}}{\cosh^2 \chi}
|0\rangle\langle 0|\nonumber\\
& & + \frac{1}{p_{\rm post}}
\sum_{n=1}^\infty(1-(1-\eta_{\rm A})^n) \frac{\tanh^{2n} \chi}{\cosh^2
\chi} |n\rangle\langle n| \; \nonumber
\end{eqnarray}
with the post-selection probability as normalization factor 
\begin{eqnarray}
p_{\rm post} &=& \frac{d_{\rm A}}{\cosh^2 \chi}+\sum_{n=1}^\infty(1-(1-\eta_{\rm A})^n) \frac{\tanh^{2n} \chi}{\cosh^2
\chi}\\
& = & \frac{d_{\rm A}}{\cosh^2 \chi} \nonumber \\
& & +\frac{1}{\cosh^2 \chi}\left(
\frac{1}{1- \tanh^2 \chi} - \frac{1}{1-(1-\eta_{\rm A}) \tanh^2 \chi}\right)
\; . \nonumber
\end{eqnarray}
This gives us the photon number distribution of the signals which are
obtained from this {\em seed  state} by polarization rotation. From
the photon number distribution we can calculate $S_m$ by summation and
$p_{\rm exp}^{\rm signal}$, via the
photodetection formula \cite{walls94a} as,
\begin{eqnarray}
S_m &=& 1-\frac{1}{p_{\rm post} \cosh^2 \chi} \left( d_{\rm A} + \eta_{\rm A} \tanh^2 \chi \right)\\
p_{\rm exp}^{\rm signal} & = & \frac{1}{p_{\rm post} \cosh^2 \chi} \\
& & \times 
\sum_{n=1}^{\infty} \left[ 1-(1-\eta_{\rm A})^n\right] \left[1-(1-\eta_T
\eta_{\rm B})^n\right] \tanh^{2 n} \chi \nonumber\\
& = & \frac{1}{p_{\rm post} \cosh^2 \chi}  \left[ \frac{1}{1-\tanh^2 \chi} \right. \nonumber \\
& & -\frac{1}{1-(1-\eta_{\rm T} \eta_{\rm B}) \tanh^2 \chi}  -\frac{1}{1-(1-\eta_{\rm A})
\tanh^2 \chi} \nonumber\\
& & \left . + \frac{1}{1-(1-\eta_{\rm A})(1-\eta_{\rm T} \eta_{\rm B}) \tanh^2 \chi}
\right] \; . \nonumber
\end{eqnarray}
As in the case of the \WCP\ scenario, we are now in the position to
calculate the gain rate of a setup from experimental parameters.  The
simulations use experimental values for the transmission line and
detectors which are the same as in the \WCP\ case. There are two
different scenarios: Either the non-degenerate downconversion produces
photons at the same frequency, or one can use downconversion with
different frequencies such that the frequency of Alice's photon has a
wavelength convenient for detection, while the other photon's
wavelength falls into one of the three telecommunication windows for
optimal propagation along the fiber or open air. To illustrate the
calculation we assumed the situation where one mode is adapted to the
$800$ nm detectors of the British Telecom experiment, while the signal
mode is emitted in one of the four modes used already for the \WCP\
case.
The results of this hypothetical experiment is shown in
Fig.~\ref{PDCgain}.
We find an increase of the covered distance against the use of the \WCP\
source, but this happens at the expense of a lower rate per signal.

To understand the decrease of the rate, we can now bound the maximal
gain per time slot in correspondence to the calculation for weak
coherent states.  It is now convenient to introduce the expected
photon number $\mu = \sinh^2 \chi$. In the optimal case, Alice's
triggering detector is perfect ($\eta_{\rm A}=1$ and $d_{\rm A}=0$), and we
neglect the negative contribution of privacy amplification and error
correction. Then we find, again using $\eta := \eta_{\rm B} \eta_T$,
\begin{eqnarray}
S_m & = & p_{\rm post} \frac{ \mu^2}{(1+\mu)^2}\\
p_{\rm exp} & = & p_{\rm post} \frac{ \eta \mu}{1+\eta \mu}
\end{eqnarray}
so that we find for the gain
\begin{equation}
G \leq \frac{1}{2} \mu \left( \frac{\eta}{1+\eta \mu} -
\frac{\mu}{(1+\mu)^2}\right)\; .
\end{equation}
Now the optimal mean photon number $\mu_{\rm opt}$ satisfies
\begin{equation}
-2 \mu_{\rm opt} - 2 \eta^2 \mu_{\rm opt}^3+\eta(1+3\mu_{\rm opt}-\mu_{\rm
opt}^2+\mu_{\rm opt}^3)=0
\end{equation}
which leads for small values of $\eta$ to $\mu \approx \frac{1}{2}
\eta$. In the same limit the gain rate is approximated by
\begin{equation}
G \approx \frac{1}{8} \eta^2 \; .
\end{equation}
This bounds the obtainable rate for the case that Bob's detectors are
perfect, so that $\eta \to \eta_T$. We find that here weak coherent
states have a potential gain rate per time slot which is twice as big
as the one of parametric down conversion. The reason is that the
photon number distribution for \PDC\ sources is basically thermal,
which shows a higher multi-photon contribution compared to a Poisson
distribution with the same mean photon number.  For practical realization,
however, a factor of two is not that significant, and the gap between
gain rate of secure bits with imperfect tools is still by orders of
magnitude separated from this limit. Therefore the question remains
open, which technology allows a simpler approach to higher rates.

Note that one would need to take into account the loss occurring when
Alice couples the photon for Bob into a fiber. This loss can be easily
incorporated in this calculations since the resulting photon number
distribution of the signals can be obtained using the photon count
formulas. Here, however, we do not study this additional
parameter. The corresponding formulas are given in appendix \ref{finiteetaC}.

\section{Conclusions}
\label{discussion}
In this paper I presented a security proof of  quantum cryptography
which is restricted to individual attacks. This proof takes into
account the non-ideal signal sources and detectors. Moreover, it
allows to compare the performance for different arrangements with
respect to the overall gain rate. In this sense it can help to decide
which type of source to use, for example weak coherent pulses or
downconversion, depending on the available technology and the task
fixing, for example, wavelength and distance. For existing
experiments, it allows to find the optimal mean photon number of the
source and the optimal working point for Bob's detectors. 

We found that the use of \PDC\ sources with a simple triggering
mechanism does not increase the overall rate of secure bits, but it
allows to increase the distance which can be covered by
experiments. The rate could be improved by a more sophisticated
detection mechanism, where Alice could, at least partly, determine the
number of pairs produced in a time slot. Even if this mechanism does
not work perfectly, it would improve the rate and distance. 

Our examples show that the use of \WCP\ sources gives, typically,
higher rates per time slot than the use of \PDC\ sources, as long as
the distance is not too big. I would like to point out again, that in
the end the total rate, that is the rate per time slot times the
repetition rate of the set-up, is what counts. It depends therefore on
the bottle-neck of the set-up which design can be made the fastest. 

The problem of non-ideal sources in the presence of loss is known
since 1995. There have been proposals to use strong reference pulses
in the two-state protocol \cite{bennett92a} and the BB84 protocol
\cite{huttner95a}, but so far these ideas have not been
implemented. The reference pulses make it more difficult for Eve to
block signals, since in those schemes Bob measures the interference of
the strong reference pulse with the weak signal, so that the absence
of the weak signal will lead to an error in half of the cases. I would
like to point out, that the security of this scheme has not been fully
analyzed yet even for individual attacks, but this scheme is certainly
the hope for the future to improve the here analyzed BB84 protocol.

\acknowledgements
 I would like to thank Mohamed Bourennane, Gilles Brassard, Mila Du\v{s}ek, Nicolas
Gisin, Richard Hughes, Bruno Huttner, Hitoshi Inamori, 
Anders Karlsson, Tal Mor, and Paul
Townsend for many discussions on the issue of security of realistic
quantum key distribution.  Furthermore, I took benefit from the 1998
quantum information workshops at ISI (Italy) and the Benasque Center
for Physics (Spain) and wish to thank their organizers and
Elsag-Bailey for support. This work has been supported by the project
 43336 of the Academy
of Finland and by the European Science Foundation (QIT programme).

\appendix
\section{Photon number splitting}
\label{photonsplitting}
The photon number splitting idea has been presented  already in
\cite{brassardsub99a}. Here I want to provide more details. 
To perform photon number splitting, Eve performs a quantum non-demolition
measurement on the total photon number in both polarization modes. As a
result the signal is now described by a $n$-photon state in the unknown
 and undisturbed signal polarization, and the photon number $n$ is known to Eve. 

The task is now to find a unitary transformation $U_{\rm PNS}^{(n)}$,
which depends on the value of $n$, such that precisely one photon from the two
signal polarization modes $a_i$ is transferred to two additional
polarization modes $b_i$ which are in Eve's hand. The polarization of
either part should be equal to the original one. This means we
require that the two signals of the first polarization basis ($+$)
transform as 
\begin{eqnarray}
\label{trafopart1}
U_{\rm PNS}^{(n)} |n,0,0,0\rangle_+ = |n-1,0,1,0\rangle_+ \\
U_{\rm PNS}^{(n)} |0,n,0,0\rangle_+ = |0,n-1,0,1\rangle_+ \; . \nonumber
\end{eqnarray}
Here the components of the state vector $|\dots\rangle_+$ correspond to the photon number
occupation of the modes $a_1, a_2, b_1, b_2$ respectively.
The requirement for the two signal states of the second polarization
basis $(\times)$ is easily formulated if we choose the mode representation defined
by the operators  $a_\pm =1/\sqrt{2}(a_1 \pm a_2)$ and $b_\pm =
1/\sqrt{2}(b_1 \pm b_2)$. The state vector
$|\dots\rangle_\times$ now denotes the occupation number in the modes
$a_+, a_-, b_+, b_-$. We require, that
\begin{eqnarray}
\label{trafopart2}
U_{\rm PNS}^{(n)}|n,0,0,0\rangle_\times = |n-1,0,1,0\rangle_\times \\
U_{\rm PNS}^{(n)}|0,n,0,0\rangle_\times = |0,n-1,0,1\rangle_\times \; .\nonumber
\end{eqnarray}

Indeed, a transformation $U_{\rm PNS}^{(n)}$ with these properties can be
found \cite{moelmer}. Eve uses an interaction described by a  Jaynes-Cummings
Hamiltonian $$H_{JC}^{(1)} = \lambda (a^\dagger_1 \sigma_1 +
a_1 \sigma_1^\dagger + a^\dagger_2 \sigma_2 +
a_2 \sigma_2^\dagger)$$ to connect the signal modes to
a three level system with one ground state $|g\rangle$  and two upper
states $|e_i\rangle$ with atomic
excitation operators $\sigma_i^\dagger$ ($i = 1,2$)
\cite{moelmer}. (For a review of the Jaynes-Cummings model see \cite{shore93a}.) The system is initially prepared in the ground state. 
After an interaction time $t = \frac{\pi}{2 \sqrt{n}
\lambda}$, which depends on $n$, the first two signal states transform
into
$|n,0\rangle_+ |g\rangle \to |n-1,0\rangle_+ |e_1\rangle$ and 
$|0,n\rangle_+ |g\rangle \to |0,n-1\rangle_+ |e_2\rangle $.
The same  dynamics involving 
two additional photonic modes, $b_1$ and
$b_2$, and the Hamiltonian 
$$H_{JC}^{(2)} = \lambda (b^\dagger_1 \sigma_1 +
b_1 \sigma_1^\dagger + b^\dagger_2 \sigma_2 + b_2 \sigma_2^\dagger)$$
 transfers (after interaction time
$\tilde{t} = \frac{\pi}{2 \lambda}$)
the excitation to
a photon in the original polarization into the modes $b_i$. In total
we have then achieved the transformations (\ref{trafopart1}) while the three-level system factors
out. 
As shown, this mechanism works fine for the first two signal
states. To see that it works for the other states as well  note that
we can introduce  a new description of the three level system with the
superpositions of the upper levels as new excited states so that
$\sigma_\pm =
1/\sqrt{2}(\sigma_1 \pm \sigma_2)$ are the new atomic operators. Then we find that the Hamiltonians,
written with these new atomic operators and with the photonic
operators in the base $(\times)$,  have the form
$H_{JC}^{(1)} = \lambda (a^\dagger_+ \sigma_+ +
a_+ \sigma_+^\dagger + a^\dagger_- \sigma_- +
a_- \sigma_-^\dagger)$ and 
$H_{JC}^{(2)} = \lambda (b^\dagger_+ \sigma_+ +
b_+ \sigma_+^\dagger + b^\dagger_- \sigma_- + b_- \sigma_-^\dagger)$.
We see, the Hamiltonians are form invariant under the the above
transformations, and it follows that this scheme performs the mapping
of (\ref{trafopart2}) as well. In general, this scheme is able to
split one photon off any $n$-photon state with definite polarization,
regardless what this polarization may be.

\section{\PDC\ with finite coupling efficiency}
\label{finiteetaC}
In this appendix I provide the straightforward derived formulas for
the case where we use a parametric downconversion source for the
triggering of the signal, and the signal travelling to Bob couples
only with a finite efficiency $\eta_{\rm C}$ into the fiber. All losses on
Alice's side which cannot be accessed by Eve can be incorporated into
this efficiency. Conditioned on a click in Alice's triggering detector
we find the following results:
\begin{eqnarray}
p_{\rm post} & = &\frac{d_{\rm A}}{\cosh^2 \chi}  \\
& & +\frac{1}{\cosh^2 \chi}\left(
\frac{1}{1- \tanh^2 \chi} - \frac{1}{1-(1-\eta_{\rm A}) \tanh^2
\chi}\right) \nonumber
\end{eqnarray}
\begin{eqnarray}
S_0 & = & \frac{1}{p_{\rm post} \cosh^2 \chi} \left(d_{\rm A} +
\frac{1}{1-(1-\eta_{\rm C}) \tanh^2 \chi} \right. \\
& & \left . -\frac{1}{1-(1-\eta_{\rm C})(1-\eta_{\rm A})
\tanh^2 \chi}\right) \nonumber
\end{eqnarray}
\begin{eqnarray}
\lefteqn{S_1  =  \frac{\eta_{\rm C} \tanh^2 \chi}{p_{\rm post} \cosh^2
\chi}} & &  \\
& & \times 
   \left(\frac{1}{1-(1-\eta_{\rm C}) \tanh^2
\chi}-\frac{1-\eta_{\rm A}}{1-(1-\eta_{\rm C})(1-\eta_{\rm A})\tanh^2\chi}\right) \nonumber\\
\end{eqnarray}
\begin{eqnarray}
S_M & = & 1-S_0-S_1\\
p_{\rm exp} & = & \frac{1}{p_{\rm post} \cosh^2 \chi}  \left[ \frac{1}{1-\tanh^2 \chi} \right.  \\
& & -\frac{1}{1-(1-\eta_{\rm T} \eta_{\rm C} \eta_{\rm B}) \tanh^2 \chi}  -\frac{1}{1-(1-\eta_{\rm A})
\tanh^2 \chi} \nonumber\\
& & \left . + \frac{1}{1-(1-\eta_{\rm A})(1-\eta_{\rm T} \eta_{\rm C} \eta_{\rm B}) \tanh^2 \chi}
\right] \; . \nonumber
\end{eqnarray}
With these quantities  we can, as before, determine the optimal gain
for a given setup.
 

\end{document}